\documentclass[12pt]{article}
\usepackage{amsmath}
\usepackage{epsfig}
\begin{document}
\begin{center}
\LARGE
\textbf{Hardy's Setup and Elements of
Reality}\\[1cm]
\large
\textbf{Louis Marchildon}\\[0.5cm]
\normalsize
D\'{e}partement de physique,
Universit\'{e} du Qu\'{e}bec,\\
Trois-Rivi\`{e}res, Qc.\ Canada G9A 5H7\\
email: marchild$\hspace{0.3em}a\hspace{-0.8em}
\bigcirc$uqtr.ca\\
\end{center}
\bigskip
\begin{abstract}
Several arguments have been proposed some years
ago, attempting to prove the impossibility of
defining Lorentz-invariant elements of reality.
Here I revisit that question,
and bring a number of additional considerations to it.
I will first analyze Hardy's argument,
which was meant to show that Lorentz-invariant
elements of reality are indeed inconsistent
with quantum mechanics.  I will then
investigate to what extent the light cone associated
with an event can be used to define Lorentz-invariant
elements of reality.  It turns out to be possible,
but elements of reality associated
with a product of two commuting operators will not always
be equal to the product of elements of reality associated
with each operator.  I will
finally examine a number of ways in which
the paradoxical features of Hardy's experiment
can be better understood.
\end{abstract}
\section{Introduction}
The notion of `element of reality' was introduced
in the famous Einstein, Podolsky and Rosen (EPR)
paper~\cite{einstein}, as an attribute of a physical
quantity whose value can be predicted with certainty
without disturbing the system.  To avoid the ambiguity
of the phrase `without disturbing the system,'
Redhead~\cite{redhead} later gave the following sufficient
condition for the existence of an element of reality,
hereafter called ER1:
\begin{quote}
If we can predict with certainty, or at any rate with
probability one, the result of measuring a physical
quantity at time $t$, then at the time $t$ there
exists an element of reality corresponding
to the physical quantity and having a value equal to
the predicted measurement result. [\textbf{ER1}]
\end{quote}

Several interpretations of quantum mechanics
involving various kinds of elements of reality were
proposed after the EPR paper, Bohmian mechanics
in particular~\cite{bohm}.  Originally developed
as nonrelativistic theories, they have been
notoriously difficult to reconcile with the
special theory of relativity.  Eventually, the question
was raised whether Lorentz-invariant elements of
reality are inconsistent with quantum
mechanics~\cite{hardy1,clifton1,clifton2}.

In this paper I will first analyze Hardy's argument,
which was meant to show that Lorentz-invariant
elements of reality are indeed inconsistent with quantum
mechanics.  I will then investigate to
what extent the light cone associated with an event
can be used to define Lorentz-invariant elements
of reality.  It turns out to be possible, but
these elements of reality won't satisfy
the so-called product rule, i.e. an element of
reality associated with a product of two commuting
operators will not always be equal to the product
of elements of reality associated with each
operator~\cite{vaidman1,vaidman2}.  I will
finally examine how, in several interpretations
of quantum mechanics, the paradoxical features of
Hardy's experiment can be better understood.
\section{Hardy's argument}
\begin{figure}[hbt]
\begin{center}
\epsfig{file=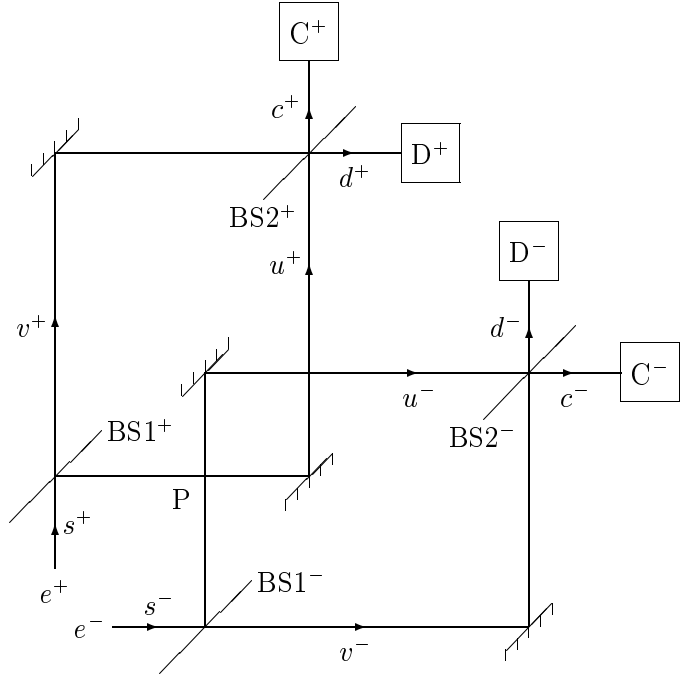,height=90mm,width=90mm}
\end{center}
\caption{Hardy's thought experiment with two
Mach-Zehnder-type interferometers}
\label{fig1}
\end{figure}

Hardy's gedanken experiment~\cite{hardy1}
is illustrated in Fig.~\ref{fig1}.
Two Mach-Zehnder-type interferometers are set up,
one for electrons (MZ$^-$, lower right) and one for
positrons (MZ$^+$, upper left).  Electron (positron)
states are prepared with initial state vectors
$|s^-\rangle$ ($|s^+\rangle$), which are wave
packets concentrated around paths $s^-$ ($s^+$)
indicated in Fig.~\ref{fig1}.  There is
annihilation with unit probability if the electron
and positron wave packets meet at point~P.

Beam splitters BS1$^{\pm}$ and BS2$^{\pm}$
act so that
\begin{align}
|s^{\pm} \rangle & \rightarrow \frac{1}{\sqrt{2}}
(i |u^{\pm} \rangle + |v^{\pm} \rangle) , \label{kets} \\
|u^{\pm} \rangle & \rightarrow \frac{1}{\sqrt{2}}
(|c^{\pm} \rangle + i |d^{\pm} \rangle) , \label{ketu} \\
|v^{\pm} \rangle & \rightarrow \frac{1}{\sqrt{2}}
(i |c^{\pm} \rangle + |d^{\pm} \rangle) . \label{ketv}
\end{align}
State vectors $|u^{\pm} \rangle$, $|v^{\pm} \rangle$,
$|c^{\pm} \rangle$ and $|d^{\pm} \rangle$ are wave
packets concentrated around associated paths in
Fig.~\ref{fig1}.

The evolution of the joint particles' state
vector is explained in Ref.~\cite{hardy1},
from which we quote the following results.
Henceforth $|\gamma\rangle$ represents photon states
resulting from electron-positron annihilation.

In any run of the experiment, there is a Lorentz
frame $F^-$ in which the electron wave packet has gone
through BS2$^-$ before the positron wave packet has reached
BS2$^+$.  During that time, the state
vector is given by:
\begin{equation}
\frac{1}{2 \sqrt{2}} (\mbox{} - \sqrt{2} |\gamma\rangle
- |u^{+} \rangle |c^{-} \rangle
+ 2 i |v^{+} \rangle |c^{-} \rangle
+ i |u^{+} \rangle |d^{-} \rangle ) .
\label{Fminus}
\end{equation}
Similarly, there is a Lorentz frame $F^+$ in which
the positron wave packet has gone through BS2$^+$
before the electron wave packet has
reached BS2$^-$.  During that time, the state
vector is given by:
\begin{equation}
\frac{1}{2 \sqrt{2}} (\mbox{} - \sqrt{2} |\gamma\rangle
- |c^{+} \rangle |u^{-} \rangle
+ 2 i |c^{+} \rangle |v^{-} \rangle
+ i |d^{+} \rangle |u^{-} \rangle ) .
\label{Fplus}
\end{equation}
Furthermore, before the positron wave packet has gone
through BS2$^+$ and the electron wave packet
has gone through BS2$^-$, the state vector
in all frames is given by:
\begin{equation}
\frac{1}{2} (\mbox{} - |\gamma\rangle
+ i |u^{+} \rangle |v^{-} \rangle
+ i |v^{+} \rangle |u^{-} \rangle
+ |v^{+} \rangle |v^{-} \rangle ) .
\label{before}
\end{equation}
Finally, when both the electron and the
positron wave packets
have gone through the second beam splitters,
the state vector in all frames is given by
\begin{align}
& \frac{1}{4} (\mbox{} - 2 |\gamma\rangle
- 3 |c^{+} \rangle |c^{-} \rangle
+ i |c^{+} \rangle |d^{-} \rangle \notag \\
& \qquad \mbox{} + i |d^{+} \rangle |c^{-} \rangle
- |d^{+} \rangle |d^{-} \rangle ) .
\label{after}
\end{align}

To argue against relativistic elements of reality,
Hardy proposes a sufficient condition for their
existence and a necessary condition for their Lorentz
invariance.  The sufficient condition essentially
coincides with Redhead's ER1.  The necessary
condition, hereafter called LI1, simply reads as:
\begin{quote}
The value of an element of reality corresponding to a
Lorentz-invariant observable is itself Lorentz
invariant. [\textbf{LI1}]
\end{quote}
I shall denote an element of reality associated
with an observable $A$ by $f(A)$.  Whenever
ER1 is satisfied for $A$, then $f(A)$ coincides
with an eigenvalue of $A$, a real number.

Suppose that in frame $F^-$, an electron is
detected in $D^-$.  From Eq.~(\ref{Fminus}), one
can predict with certainty that a measurement of the
observable $U^+ = |u^+\rangle \langle u^+|$ will
yield the value~1.  Hence $U^+$ is an element of
reality, and $f(U^+) = 1$.  Likewise suppose that
in frame $F^+$, a positron is detected in $D^+$.
From Eq.~(\ref{Fplus}), this implies that
$U^- = |u^- \rangle \langle u^- |$ is an
element of reality, and $f(U^-) = 1$.  According
to Eq.~(\ref{after}), both these situations
will occur together, on average, in one of every
sixteen runs.  In any such case, LI1 implies that
both $U^+$ and $U^-$ are
elements of reality.  But Hardy claims that
\begin{equation}
f(U^+) f(U^-) = 1 \Rightarrow f(U^+ U^-) = 1 .
\label{implic}
\end{equation}
Hence in every run where an electron is detected
in $D^-$ and a positron is detected in $D^+$,
we obtain that $f(U^+ U^-) = 1$.

From Eq.~(\ref{before}), however, we can predict
with certainty that a measurement of the observable
$U^+ U^-$ will yield the value~0.  Thus
$f(U^+ U^-) = 0$, which contradicts the result
of the previous paragraph.

The upshot is that there seems to be no way to
assign elements of reality in a relativistically
invariant way.
\section{Analysis}
I have argued elsewhere~\cite{marchildon1} that
the inference made in~(\ref{implic}) assumes that
ER1 is not only a sufficient, but also a necessary
condition.  This conclusion is reinforced by the
fact that variants of Hardy's argument~\cite{clifton2}
assume the validity of the product
rule for commuting operators:
\begin{equation}
f(U^+) f(U^-) = f(U^+ U^-) .
\label{product}
\end{equation}
Indeed one can show~\cite{marchildon1}
that any real-valued function~$f$ which (i) is
defined on a maximal set of commuting Hermitian
operators, (ii) satisfies Eq.~(\ref{product}),
and (iii) is~1 on some but not all one-dimensional
projectors in the set, singles out a one-dimensional
subspace of the state space, i.e.\ there is a unique
one-dimensional projector in the set on which $f = 1$.
If the element of reality that this function assigns
is identified with an eigenvalue of a quantum
observable, then a unique state
vector is singled out by the specification that $f = 1$.
Hence it leads to the most definite predictions
that quantum mechanics allows.

Since the contradiction obtained in Sec.~2
involves somewhat more than the sufficient
condition~ER1 and the necessary condition~LI1,
one can ask whether it is indeed possible
to have Lorentz-invariant elements of reality.

To answer this question, let us first note that
condition ER1 involves the word `predict'
in an essential way.  Since `predict'
refers to the future, i.e. to times later
than some instant $t$ in a given Lorentz frame,
ER1 is clearly not a relativistically invariant
criterion.  To get an invariant criterion, we should
introduce an invariant specification, i.e.\ we
should make use of the light cone.  I have shown
in Ref.~\cite{marchildon1} that the backward light
cone is not an appropriate choice, because it
does not capture the kind of elements of reality
that EPR had in mind.  But the forward
light cone is.  So here's what a relativistically
invariant sufficient condition for the existence
of an element of reality may look like:
\begin{quote}
If from the (relevant) information on or outside
the forward light cone of a possible measurement event
$\mathcal{E}$, we can infer with certainty,
or at any rate with probability one, the result
of measuring a physical quantity at
$\mathcal{E}$, then at that event there
exists an element of reality corresponding
to the physical quantity and having a value equal to
the predicted measurement result. [\textbf{ER3}]
\end{quote}

Criterion ER3 bears some relation to the
Hellwig-Kraus approach~\cite{hellwig} to state vector
collapse, of which I'll say more in the next section.
In that approach, the collapse occurs on the backward
light cone of the measurement event.  It therefore
allows for relevant information obtained through
collapse to bear upon a measurement performed in its
relative (though not absolute) past, just like ER3
does.

\begin{figure}[hbt]
\begin{center}
\epsfig{file=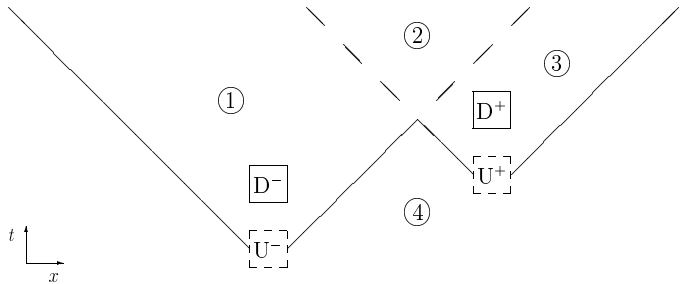,height=37.3mm,width=90mm}
\end{center}
\caption{Combination of two light cones}
\label{fig2}
\end{figure}

To apply to Hardy's argument, criterion ER3 should
also be adapted to nonlocal observables like
$U^+ U^-$.  How can this be done? Fig.~\ref{fig2}
shows two light cones associated with $U^+$ and
$U^-$ respectively, or more precisely with
elements of reality pertaining to these
observables at the events in dashed boxes.
We can combine the exteriors of the two light
cones through their union $\mathcal{U}$ (involving
regions~1, 3 and~4) or their intersection $\mathcal{I}$
(involving only region~4).

Suppose that we try to adapt criterion ER3 to
$\mathcal{U}$.  It is clear
that any time an actual measurement of $U^+ U^-$ is
performed, its result can be deduced from information
available in $\mathcal{U}$, for this
region includes part of the absolute future of the
measurement events.  Hence the criterion is in this
case trivial.  (We'll turn later to the case where the
measurement is not performed.)

Suppose next that we try to adapt criterion ER3
to~$\mathcal{I}$.  Consider as we did before a run
where detection occurs at $D^+$ and $D^-$.
Since $D^-$ is outside the forward light cone of
$U^+$, ER3 implies that $f(U^+) = 1$.  Likewise
since $D^+$ is outside the forward light cone of
$U^-$, ER3 implies that $f(U^-) = 1$.  This holds
in all Lorentz frames.

It turns out, however, that neither $D^+$ nor $D^-$
are in~$\mathcal{I}$, the intersection of the
regions outside the two forward light cones.
Hence ER3 cannot be used to deduce the existence
of an element of reality associated with $U^+ U^-$,
nor \emph{a fortiori} to attribute a value to
$f(U^+ U^-)$.  Hardy's argument therefore no longer
goes through, and the product rule~(\ref{product}) no
longer necessarily holds.

We should note that criterion ER3 for the
existence of elements of reality, and its
generalization to nonlocal observables through
the region $\mathcal{I}$, involve
context~\cite{clifton1,berndl}.
Indeed the `relevant information' is not the
same for the local observable $U^+$ as it is for
the nonlocal observable $U^+ U^-$.

Vaidman~\cite{vaidman1,vaidman2} has analyzed
Hardy's setup using the Aharonov, Bergmann and Lebowitz
rule~\cite{ABL}, which evaluates the probability of
measurement results conditional on postselection
as well as preselection.  He then found that there
is unit probability for the following three
intermediate (unperformed) results:
(i) $U^+$ yields 1; (ii) $U^-$ yields 1; and (iii)
$U^+ U^-$ yields 0.  In this sense the product rule
for elements of reality does not hold.

In Vaidman's analysis, the value of the element
of reality corresponding to $U^+ U^-$ can be
deduced from information in the union
$\mathcal{U}$ of the regions outside the two
forward light cones of the events.  As he claims,
it thus provides Lorentz-invariant elements
of reality.
\section{Understanding Hardy's experiment}
Hardy's thought experiment, like many others in quantum
mechanics, shows paradoxical features related
to correlations over spacelike separations.
I have argued elsewhere~\cite{marchildon2},
following others~\cite{fraassen}, that
interpreting quantum mechanics means answering the
question, How can the world be for the theory
to be true?  To attenuate, if not resolve, the
paradoxical features of Hardy's setup, let us try to
understand it in four different interpretations.
\subsection{Collapse theories}
Hardy originally framed his argument against
Lorentz-invariant elements of reality in the
language of state vector collapse.  The argument,
however, does not depend on that assumption.
Nevertheless, collapse theories do provide a
way to understand what happens in the experiment.

Assume as before that detectors $D^+$ and $D^-$
eventually fire, at instants we shall denote
by $t_+$ and $t_-$.  For simplicity, we also
assume here that detectors $D^{\pm}$ and $C^{\pm}$
are very close to the second beam splitters, so
that the time interval between splitting and
detection is negligible.

In a nonrelativistic collapse
theory such as von Neumann's~\cite{neumann},
collapse occurs on an equal time hypersurface.
Clearly, this singles out a preferred reference
frame, since equal time in one Lorentz frame
is not equal time in others.

Let us now consider two different cases.  In the
first one, the preferred frame is the one where
the two detections are simultaneous.
Then for $t < t_+ = t_-$, the state vector
$|\psi\rangle$ is given by Eq.~(\ref{before}),
whereas for $t > t_+ = t_-$, it is given by 
$|d^+\rangle |d^-\rangle$.  In the second
case, the preferred frame is one where
detection at $D^+$ occurs before detection
at $D^-$, i.e.\ $t_+ < t_-$.  Then we have
\begin{equation}
|\psi\rangle = \begin{cases}
\mbox{Eq. } (\ref{before}) & \mbox{if } t < t_+ , \\
|d^+\rangle |u^-\rangle & \mbox{if } t_+ < t < t_- , \\
|d^+\rangle |d^-\rangle
& \mbox{if } t_- < t . \end{cases}
\end{equation}
We can see that the observable $U^-$ is an element
of reality in the second case, but not in the first
one.  This just illustrates the lack of Lorentz
invariance of nonrelativistic collapse.

\begin{figure}[hbt]
\begin{center}
\epsfig{file=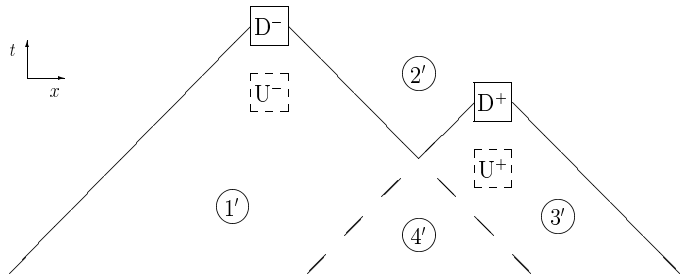,height=36mm,width=90mm}
\end{center}
\caption{Collapse in the Hellwig-Kraus theory}
\label{fig3}
\end{figure}

The relativistic collapse theory of Hellwig and
Krauss~\cite{hellwig}, applied to the present
situation, is illustrated in Fig.~\ref{fig3}.
In each of the four regions, the state
vector is given as follows:
\begin{equation}
|\psi\rangle = \begin{cases}
|d^+\rangle |u^-\rangle & \mbox{in }1' , \\
|d^+\rangle |d^-\rangle & \mbox{in }2' , \\ 
|u^+\rangle |d^-\rangle & \mbox{in }3' , \\
\mbox{Eq. } (\ref{before}) & \mbox{in }4' . \end{cases}
\end{equation}
One can see that $U^-$ is an element of reality
in region~$1'$, $U^+$ is an element of reality
in~$3'$ and $U^+ U^-$ is an element of reality
in~$4'$.  These regions being different, the
product rule doesn't hold.
\subsection{Bohmian mechanics}
In Bohmian mechanics, positions of particles
are elements of reality.  The electron and the
positron both follow deterministic trajectories
in one arm of their corresponding interferometers.

The trajectories, however, are not
relativistically covariant.  There is a preferred
frame where they are to be computed.
If the preferred frame is the one where both
measurements are simultaneous, and detectors
$D^+$ and $D^-$ fire, the trajectories
can consistently avoid the $u^+ u^-$ path.
If, on the other hand, the preferred frame is the
one where the positron is measured first, then
the state vector before the positron measurement
is given by Eq.~(\ref{before}).
After measurement, it effectively
becomes $|d^+\rangle |u^-\rangle$.  The electron
has gone through $u^-$, but one cannot say (as would
presumably be found upon explicit calculation)
that the positron has gone through~$u^+$.

Bohmian elements of reality are Lorentz invariant, but
condition ER1 is not in general valid.  That is,
prediction with certainty in one frame is not enough to
ascertain the existence of an element of reality.

Note that in Bohmian mechanics, although
particle trajectories are not relativistically
covariant, statistical predictions are, since
they coincide with the ones made by standard
quantum mechanics.
\subsection{Everett's relative states}
In Everett's relative-states (or many-worlds)
theory~\cite{everett}, the state vector never collapses.
All components of the final state
vector~(\ref{after}) coexist.  Different variants
of Everett's approach will take the coexistence
to apply to different worlds, different
minds or different decohering sectors of the
state vector.

In a world (say) associated with the
$|d^+\rangle |d^-\rangle$ component of state
vector~(\ref{after}), detectors
$D^+$ and $D^-$ fire.  Elements of reality can
be associated with $|d^+\rangle \langle d^+|$ and
$|d^-\rangle \langle d^-|$ only in such worlds.
In other worlds just as real as these, however,
detectors $D^+$ and $C^-$ fire.  Elements of reality
can there be associated with $|d^+\rangle \langle d^+|$
and $|c^-\rangle \langle c^-|$.

In Everett's theory, ER1 is not a sufficient
condition for the existence of an element of
reality.  If $D^-$ fires in frame $F^-$, state
vector~(\ref{Fminus}) cannot in general be used
to attribute an element of reality to
$|u^+\rangle \langle u^+|$.  To do so, one
would need to put detectors in paths $v^+$ and
$u^+$, in addition to detectors $D^-$ and $C^-$.
In that case the various terms of~(\ref{Fminus})
would correspond to different worlds.  In all
worlds where $D^-$ would fire, the detector
in $u^+$ would too.

Everett's theory is relativistically covariant.
In any reference frame,
the positron observer and the electron observer,
if they eventually come close along a timelike
path, will always find themselves with the proper
correlations.
\subsection{Cramer's transactional view}
In Cramer's transactional interpretation~\cite{cramer},
the electron-positron source emits a (retarded)
offer wave while various detectors (including the
gamma-ray detectors registering electron-positron
annihilation) respond with (advanced)
confirmation waves.  This is reminiscent of
Wheeler-Feynman electrodynamics.

Eventually a transaction is established between
the source and one detector for each particle.
The transaction involves the actual (as opposed
to counterfactual) configuration of the
measuring devices.  Counterfactual reasoning
cannot lead to an inference of elements of
reality.

The process is relativistically covariant.
No element of reality is attached to a trajectory
or partial state vector independently of
a transaction.
\section{Conclusion}
In quantum mechanics, elements of reality are
not easy to reconcile with Lorentz invariance.
We have seen that they can if the product rule
is abandoned.  Different interpretations
of quantum mechanics may or may not assign
elements of reality, and if they do they may
do so differently.  It is the present author's
contention, however, that the existence of
various possible ways to do so illuminates
our understanding of quantum mechanics.
\section*{Acknowledgements}
I am thankful to an anonymous referee for
comments helping to make the argument clearer.
This work was supported by the Natural Sciences and
Engineering Research Council of Canada.
\end{document}